\documentclass[11pt,oneside,a4paper]{article}
\usepackage[utf8]{inputenc}
\usepackage{cite}
\usepackage{amsmath,amssymb,amsfonts}
\usepackage{tabularx}
\usepackage{graphicx}
\usepackage{textcomp}
\usepackage{threeparttable}
\usepackage{tabulary}
\usepackage{color,soul}
\usepackage{authblk}
\usepackage[margin=0.5in]{geometry}
\usepackage{subcaption}
\usepackage{url}
\usepackage{lineno}
\usepackage{float}
\usepackage{soul}
\usepackage{multirow}
\usepackage{tabularx}
\usepackage{rotating}
\usepackage{algpseudocode}
\usepackage{algorithm}
\usepackage[titletoc]{appendix}
\usepackage{mathtools}
\usepackage{amsmath}
\usepackage{rotating}
\usepackage[export]{adjustbox}
\usepackage{graphicx}
\usepackage{caption}
\usepackage{subcaption}
\usepackage{color,soul}
\usepackage{amsmath}
\usepackage{tabularx}
\usepackage{xltabular}

\usepackage{url}

\title{An ontological approach to foster the convergence, interoperability and operationalization of frameworks for Trustworthy AI}

\author{Salvatore F. Pileggi \thanks{SalvatoreFlavio.Pileggi@uts.edu.au}}

\affil{School of Computer Science, Faculty of Engineering and IT, University of Technology Sydney, Australia}

\begin{document}

\begin{titlepage}
\maketitle
\end{titlepage}
 
\begin{abstract}
AI systems are consistently evolving in terms of both capability and autonomy with an holistic social impact. In this context of proliferation and fast technological evolution, the scientific community is actively engaged to assure Trustworthy AI. However, in general terms, AI safety research is significantly slower and is facing critical challenges in terms of strategy, consensus and operationalisation. 
This paper presents AI-Ethics Ontology (AI-EO) which, by leveraging Semantic Technologies on the Web infrastructure and ontology-based knowledge representations, provides an abstracted semantic infrastructure to foster the convergence, interoperability and operationalization of the different frameworks for Trustworthy AI. The current implementation results from the analysis of two relevant case studies to establish a dynamic development process in fact, as well as to enable its iterative evolution according to a formally-defined methodology. The version 1.0 of the Ontology is freely available and has been designed to be conceptually close to target applications, in a context of interoperability, adaptability as a natural response to change and usability.
\end{abstract}

\textbf{Keywords:} AI Ethics, Trustworthy AI, Socially-responsible AI, Ontology, Semantic Web.

\section{Introduction}

AI systems are consistently evolving~\cite{shao2022tracing} in terms of capability and autonomy with a concrete and tangible impact on different aspects of the society. It applies to an individual level, with Generative AI becoming progressively part of the everyday life~\cite{fui2023generative}, as well as more holistically, for instance at an organization level, where AI is contributing to maximise value and capability across different dimensions (e.g.~\cite{mikalef2021artificial}\cite{chowdhury2023unlocking}\cite{van2023dynamics}).Additional emerging paradigms, such as Agentic AI~\cite{acharya2025agentic}, are defining a mainstream towards General AI~\cite{fei2022towards}. 

In this context of proliferation and fast technological advance, the scientific community is actively engaged to assure Trustworthy AI~\cite{li2023trustworthy}, with a priority on safety, transparency and fairness that should be guaranteed by a number of authoritative ethical principles and related practical requirements.
However, despite the important initiatives~\cite{liang2022advances}, AI safety research is significantly slower than AI evolution and is facing critical challenges in terms of strategy, consensus and operationalisation~\cite{bengio2024managing}. 

This paper addresses the potential contribution of the Semantic Web~\cite{lassila2001semantic} and its implementation over W3C standards~\cite{W3CStandards} to foster the convergence, interoperability and operationalization of frameworks for Trustworthy AI. The consolidation of the Semantic Web within the different domains~\cite{patel2021present} demonstrates the relevance and effectiveness of Ontologies~\cite{guarino2009ontology} to express formal semantics in a context of interoperability over the Web.

\paragraph{Related Work.}

As far as the author knows, AIPO~\cite{harrison2021ontology} is the first and, probably, the most comprehensive ontology in scope. Based on the OECD’s AI principles~\cite{OECD}\cite{yeung2020recommendation}, this ontological support is explicitly designed to support and facilitate the consolidation of the growing corpus of knowledge through the Semantic Web~\cite{lassila2001semantic}. 

Another interesting ontology-based conceptualization is provided by TAIR~\cite{hernandez2024ontology}, that explicitly targets the requirements for trustworthy AI. Such a Requirement Engineering perspective is gaining some popularity and is, indeed, addressed also in other works~\cite{guizzardi2023ontology}\cite{sadovski2024towards}\cite{guizzardi2024using}.

The solution proposed in~\cite{lewis2021ontology} is specifically aimed at standardization with a focus on stakeholders and organizational context, while~\cite{rismani2023does} mainly addresses roles and skills. AI Risk and its Management is object of the work presented in~\cite{golpayegani2022airo}.

Last but not least, there is some trend to a domain-specific approach. For instance, the ontological approaches in~\cite{houghtaling2023standardizing}\cite{prestes2021first} look at the formalization of standards for ethically driven robotics and automation systems.

\paragraph{Aims and Scope.} This paper presents a novel ontology, AI-Ethics Ontology (AI-EO), that is significantly different from the existing ones in the field as it aims at providing an abstracted semantic infrastructure, which is conceptually closer to target applications. By leveraging Semantic Technologies on the Web infrastructure~\cite{lassila2001semantic} and ontology-based knowledge representations~\cite{guarino2009ontology}, as extensively discussed in the paper, AI-EO is explicitly designed to: 
 
\begin{itemize}
\item foster the semantic convergence among the different frameworks and their interoperability; 
\item operationalise principles
\item enable a cross-framework approach
\item facilitate the consolidation and semantic enrichment of the resulting environments
\item enable in fact a process of evolution of the ontology as a natural response to change 
\item establish a descriptive (over prescriptive) approach to maximise adaptability and usability
\end{itemize}
 
\paragraph{Structure of the paper.} The paper follows with an overview of the research background (Section~\ref{sec:background}) and of methodological aspects (Section~\ref{sec:methodology}), while the implementation of the ontology and its discussion are object of Section~\ref{sec:implementation} and~\ref{sec:discussion} respectively.

\section{Research Background}\label{sec:background}

The ontology proposed in this paper results from the analysis of two different well-known frameworks, \textit{Australia’s AI Ethics Principles} and \textit{EU's Ethics Guidelines for Trustworthy AI}, which are briefly described in this section. 

Each framework is an input that triggers a development iteration according to the reference methodology (Section~\ref{sec:methodology}). These two iterations aim to provide a fundamental proof of concept by (\textit{i}) establishing the ontology development process and (\textit{ii}) enabling its iterative evolution. 
As discussed later on in the paper (Section~\ref{sec:discussion}), this incremental approach is preferred to a more consistent development step underlined by a larger number of input frameworks/interactions. While both frameworks are centred on ethical principles, they present some significant difference in structure and context.

\paragraph{Australia’s AI Ethics Principles.}

Australia has currently identified 8 Ethics Principles to ensure safe, secure and reliable AI in a context of risk reduction in business and governmental practices to design, develop and implement AI systems~\cite{AU_principels}. The framework is explicitly considered to be "entirely voluntary" and  based on the \textit{Ethically aligned design} report by IEEE~\cite{chatila2017ieee}.

\paragraph{Ethics Guidelines for Trustworthy AI (EU).} The EU approach to the use of AI is regulated by a specific act~\cite{EU_act}, which is considered the first actual AI law. The underlining regulatory framework~\cite{EU_trustwortyAI} is more structured to reflect a more comprehensive approach. The core part of the framework includes a set of core principles and a set of requirements, which define the foundations and the realisation of Trustworthy AI respectively. Such concepts are framed looking at three different pillars - i.e. lawful, ethical and robust AI. Additionally, they explicitly refer to fundamental human rights.

\section{Methodology and Approach}\label{sec:methodology}

The relevance of methodological aspects in ontology development has been historically recognised~\cite{jones1998methodologies} and has progressively led to the establishment of an engineering approach - i.e. Ontology Engineering (OE) - to deal with the complexity of the abstractions within the different domains~\cite{spyns2002data} assuming a networked world~\cite{suarez2011introduction}.

Many strategies and methodologies have been proposed in the last two decades. For instance, among others, human-centres~\cite{kotis2006human}, collaborative~\cite{simperl2014collaborative} and Agile~\cite{peroni2016simplified} development. OE is still evolving as a response of new needs and challenges~\cite{tudorache2020ontology}.
However, despite the resulting body of knowledge, there is currently no pre-defined and commonly accepted systematic approach to ontology development.

\begin{figure}[!h]
\centering
\begin{subfigure}{.48\textwidth}
\centering
    \includegraphics[width=.85\textwidth]{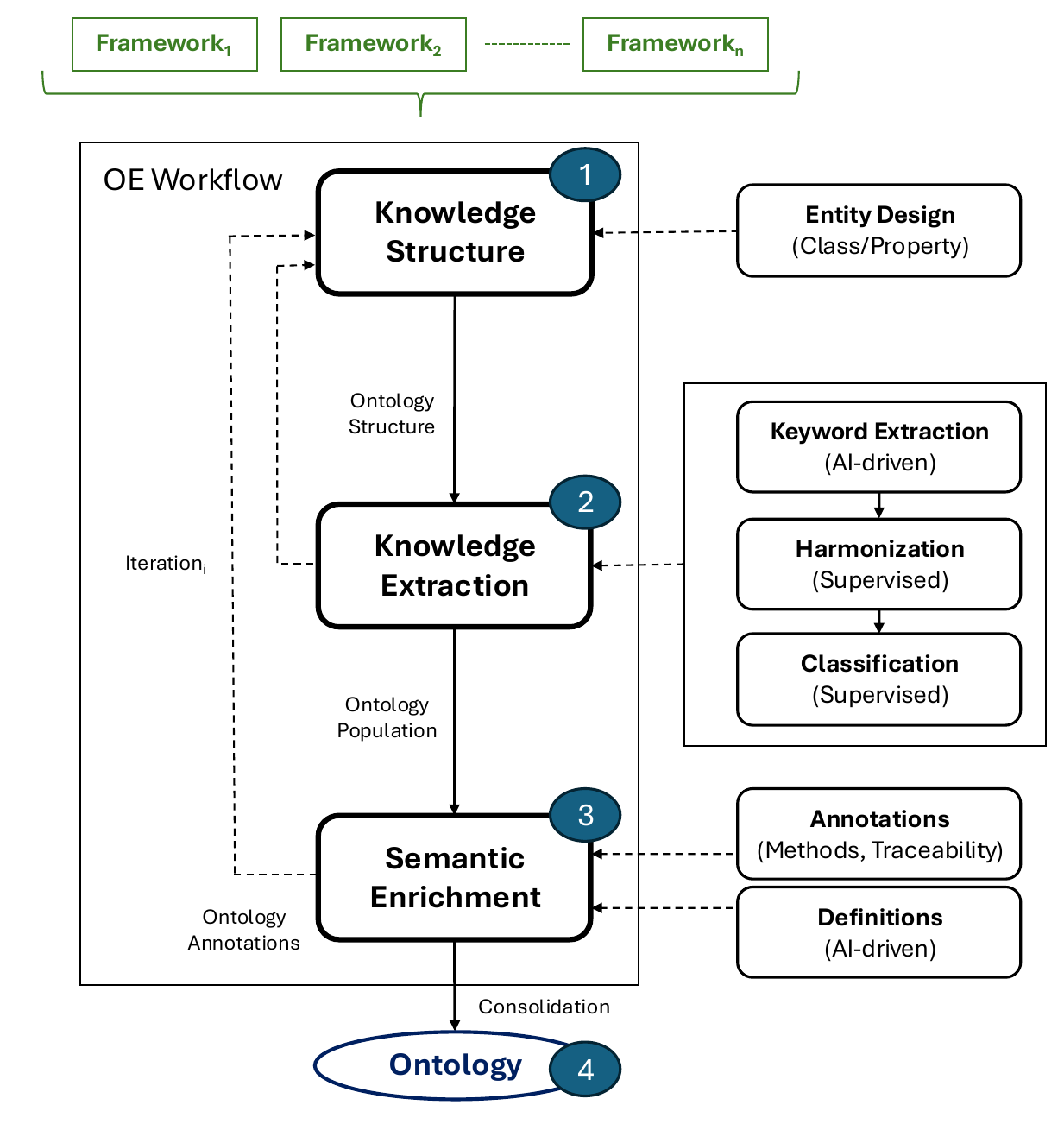}
    \caption{OE process.}\label{fig:workflow}
\end{subfigure}
~
\begin{subfigure}{.48\textwidth}
\centering
    \includegraphics[width=.85\textwidth]{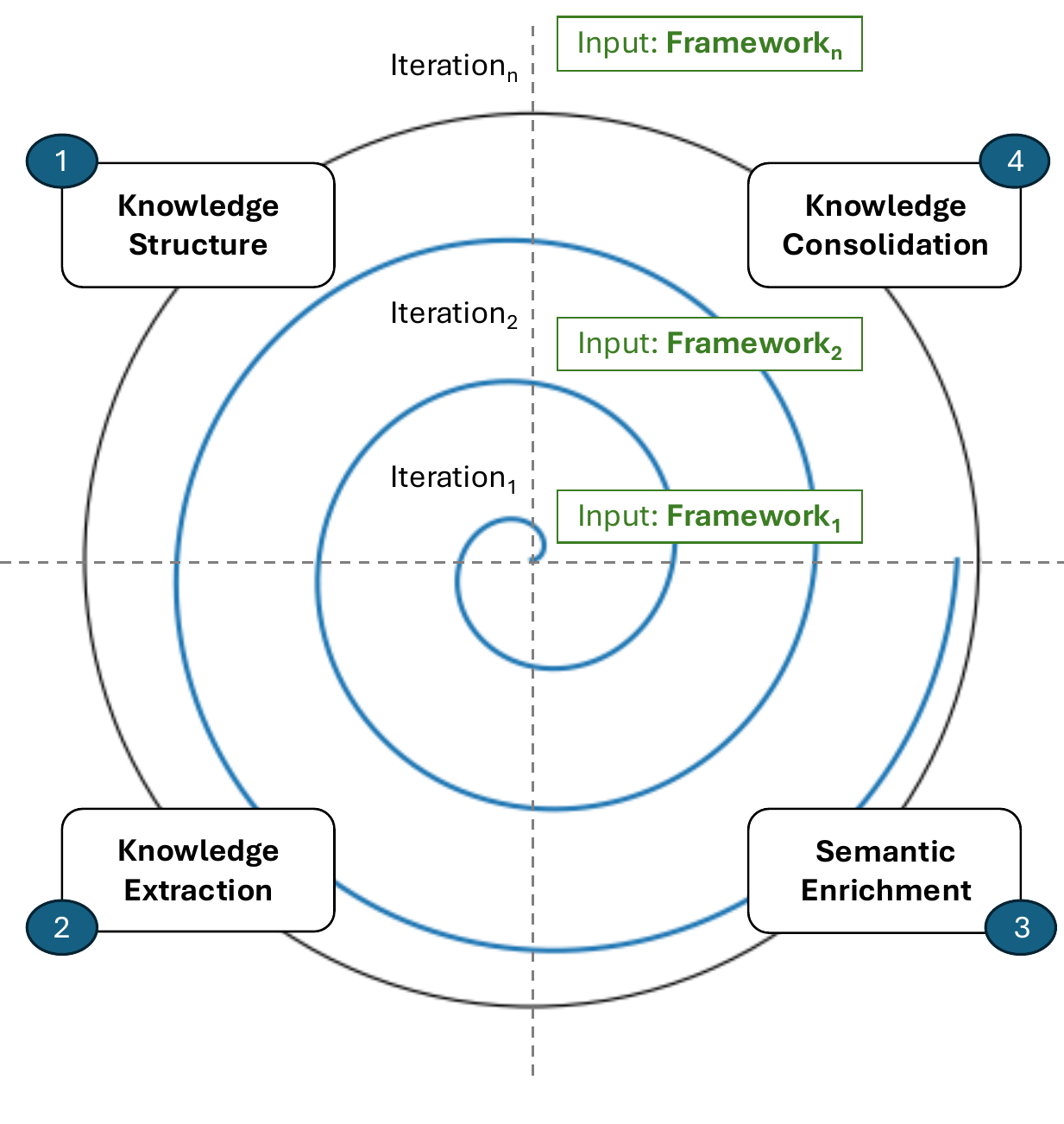}
    \caption{Iterative evolution of the Ontology.}\label{fig:spiral}
\end{subfigure}
\caption{Workflow-based representation of the Ontology Engineering (OE) process and iterative evolution of the ontology as a response of additional input.}\label{fig:KE}
\end{figure}

\paragraph{Iterative Ontology Development.}

The iterative ontology development process is represented by the workflow in Fig.~\ref{fig:workflow}. The input is a set of frameworks for trustworthy AI. An iteration, which includes the analysis of the input, as well as its context-specific conceptualization and associated ontological design,  is enabled by a single input framework. Therefore, \textit{n} input frameworks implies an equivalent number \textit{n} of iterations. The current implementation proposed later in the paper (Section~\ref{sec:implementation})  corresponds to two iterations with the frameworks previously described (Section~\ref{sec:background}) as an input. 

From an OE perspective, the process is structured in four different stages as follows:

\begin{itemize}

\item {\textit{Knowledge Structure}. The input is analysed by conceptualisation and its intrinsic complexity is converted into an ontological structure according to the W3C standards~\cite{W3CStandards}. This is a critical human-performed modelling step as it identifies the key concepts in the domain and their characterization, as well as it aims at modelling existing relationships.}

\item {\textit{Knowledge Extraction}. The unstructured input is converted into a semi-structured content aligned with the ontological schema. The extraction of the most relevant keywords is AI-driven, while their classification and harmonisation within the resulting ontology is currently supervised.}

\item {\textit{Semantic Enrichment}. Semantic annotations, including references, definitions, methods and related tools/parameters, are associated with the entities in the ontology. That's critical to establish traceability and transparency, as well as to allow the evolution of the ontology as a response to external changes.}

\item {\textit{Knowledge Consolidation}. The knowledge resulting from a given iteration is further elaborated to optimise its integration by establishing semantic equivalences and, more in general, by consolidating its value in the context of the ontology.}

\end{itemize}

\paragraph{Saturation and Convergence.}

Looking holistically at the OE process (Fig.~\ref{fig:spiral}), in a first phase the resulting knowledge expressed by the ontology is expected to significantly expand at any new iteration. This expansion phase can be reasonably be modelled as a spiral that reflects an Agile incremental approach. Such an increment may be measured in different ways by considering more or less abstracted metrics based on ontological fundamental metrics. Examples of the latter are in Table~\ref{tab:metrics}.

Despite the potential diversity in terms of structure and approach of the different frameworks in input, their common focus progressively facilitates the convergence of views towards an unified and semantically enriched framework. Such a phase of convergence is characterised by a consolidation of the knowledge and results in a saturation of the process corresponding to a smaller increment for a new iteration.

\section{AI-Ethics Ontology (AI-EO) implementation}\label{sec:implementation}

The first version of the AI-Ethics Ontology has been implemented in OWL 2~\cite{hitzler2009owl} with the support of Protege~\cite{gennari2003evolution}, a popular ontology development tool. The main statistics (as provided by such environment) are reported in Table~\ref{tab:metrics}. Additionally, a preliminary set of tests has been conducted adopting two different reasoners - i.e. Pellet~\cite{sirin2007pellet} and HermiT~\cite{glimm2014hermit}) - and related wrappers to enable complex query over ontologies in SPARQL~\cite{angles2008expressive}.  

\begin{table}[h]
\small
\centering
\caption{Main statistics on the Ontology provided by Protege.} \label{tab:metrics}

\begin{tabularx}{.4\textwidth}{  l | c  }
\hline
\hline
\textbf{Metric} &  \textbf{Value}\\
\hline
Axiom & 1097 \\
Logical axioms count & 696\\
Declaration axioms count & 224\\
Class count & 19\\
Object property count & 10\\
Data property count & 0\\
Individual count & 192\\
Annotation property count & 4\\
\hline

\hline
\end{tabularx}
\end{table}

In order to provide a comprehensive overview of the current implementation, this section is structured in three parts to concisely describe the ontological schema, the most relevant semantics supported by the ontology and a potential applications.

\subsection{Ontological Schema}

The backbone of the ontology is defined by a set of main concepts (\textit{Classes} in OWL) and by the constructs (\textit{Object Properties} in OWL) to establish relationships among class instances. Such OWL entities for the current implementation are reported in Table~\ref{tab:classes} and~\ref{tab:opjectProperties} respectively.   

\begin{table}[!h]
\small
\centering
\caption{Main Classes.} \label{tab:classes}

\begin{tabularx}{\textwidth}{  l | c | c | X }
\hline
\hline
\textbf{Class} & \textbf{Sub-class of} & \textbf{Equivalent to} & \textbf{Disjoint with}\\
\hline
\hline
\multicolumn{4}{c}{\textit{Central Concepts}}\\
\hline
\textit{AI\_Dimension} & - & - & Framework, Principle, FundamentalRight, Requirement \\
\textit{Framework} & - & - & Principle, AI\_Dimension, FundamentalRight, Requirement  \\
\textit{FundamentalRight} & - & - & Framework, Principle, AI\_Dimension, Requirement\\
\textit{Principle} & - & - & Framework, FundamentalRight, AI\_Dimension\\
\textit{Requirements} & - & - & Framework, FundamentalRight, AI\_Dimension \\
\hline
\multicolumn{4}{c}{\textit{Materialisation \& Association}}\\
\hline
\textit{Application} & - & UseCase, Scenario & -\\
\textit{Example} & - & - & -\\
\textit{Scenario} & - &  UseCase, Application & - \\
\textit{UseCase} & - & Scenario, Application & -\\
\hline
\multicolumn{4}{c}{\textit{Concept Analysis \& Classification}}\\
\hline
\textit{Keyword} & - & - & -\\
\textit{Characteristic\_keyword} & Keyword & - & -\\
\textit{Development\_keyword} & Keyword & - & -\\
\textit{EnvironmentalDimension\_keyword} & Keyword & - & -\\
\textit{GovernamentalDimension\_keyword} & Keyword & - & -\\
\textit{IndividualDimension\_keyword} & Keyword & - & -\\
\textit{OrganizationalDimension\_keyword} & Keyword & - & -\\
\textit{Risk\_keyword} & Keyword & - & -\\
\textit{SocialDimension\_keyword} & Keyword & - & -\\
\textit{SustainableDevelopment\_keyword} & Keyword & - & -\\
\hline
\hline
\end{tabularx}
\end{table}

Classes are conceptually structured in three different subsets: the key concepts, which include the most abstracted and central concepts - i.e.framework, principle, requirement, fundamental right and dimension; concepts to "materialise" the different abstractions by defining concrete examples, scenarios, use cases and applications; and, finally, concepts to support analysis and classification.     

\begin{table}[!h]
\small
\centering
\caption{Object Properties.} \label{tab:opjectProperties}

\begin{tabularx}{.85\textwidth}{  l | c | c | c | c }
\hline
\hline
\textbf{Object Property} & \textbf{Sub-property of} & \textbf{Domain} & \textbf{Range} & \textbf{Equivalent to}\\
\hline
\hline
\textit{application} & - & - & Application & scenario, useCase\\
\hline
\textit{dimension} & - & - & AI\_Dimension & - \\
\hline
\textit{example} & - & - & Example & -\\
\hline
\textit{fundamentalRight} & - & - & FundamentalRight & - \\
\hline
\textit{keyword} & - & - & Keyword & - \\
\textit{relevantKeyword} & keyword & - & Keyword & -\\
\hline
\textit{principle} & - & - &  Principle & - \\
\hline
\textit{requirement} & - & - & Requirement & - \\
\hline
\textit{scenario} & - & - & Scenario & application, useCase\\
\hline
\textit{useCase} & - & - & UseCase & application, scenario\\
\hline
\end{tabularx}
\end{table}

Complementarily, object properties are designed to link and semantically relate the different concepts for the concept instances (\textit{Individuals} in OWL) in the ontology.

\subsection{Specification of Semantics}

The semantic enrichment of the ontology is critical to proper supports the application of the ontology within the different systems. In AI-EO, the specification of semantics includes a number of different aspects as follows:

\begin{itemize}

\item \textit{Annotations} aim at specifications though labels and short descriptions, at enabling traceability by linking references and transparency by specifying methods of analysis and related parameters. The list of annotations (\textit{Annotation properties} in OWL) in the current implementation is reported in Table~\ref{tab:annotationProperties}.

\begin{table}[!h]
\small
\centering
\caption{Annotation Properties.} \label{tab:annotationProperties}

\begin{tabularx}{.85\textwidth}{  l | X  }
\hline
\hline
\textbf{Annotation Property} &  \textbf{Description/Use}\\
\hline
\hline
\textit{method} & It allows to track methods, techniques and tools. For instance, in the current implementation, is is adopted to annotate the keyword extraction tool and related parameters for the object property \textit{relevantKeyword}.\\
\hline
\textit{reference} & As per common meaning.\\
\hline
\textit{shortDescription} & It allows to add a short description/definition for an ontological resource. Ideally, it should be used in conjunction with \textit{reference}.\\
\hline
\textit{rdfs:label} & As per meaning in the original RDF Schema vocabulary~\cite{RDFS}, it is adopted to provide a human-readable version of a resource's name.\\
\hline
\end{tabularx}
\end{table}
 
\item \textit{Semantic constraints} are limited to a number of disjointness relationships involving central concepts only (Table~\ref{tab:classes}). That is a very minimal set of semantic constraints to avoid specifications which are not consistent with original definitions or potentially mis-leading. For instance, according to the current analysis, a framework includes a number of principles or requirements; defining an instance which is at the same time a framework and a principle/requirement would be not semantically consistent. 

\item \textit{Structural inference} allows to infer additional not explicitly defined semantics from the ontology structure. A given property $P$ establishes a relationship between two individuals $x$ and $y$ through a triple $\{x, \, P, \, y\}$. Inference rules on the \textit{domain} enable inference on the type of $x$, while rules on the \textit{range} enable inference on $y$. The current implementation of the ontology includes a number of inference rules on the range only (Table~\ref{tab:opjectProperties}). That is because the ontology wants to provide an open descriptive vocabulary rather than fostering a prescriptive approach.  For instance, the triple $\{fairness, \, keyword, \, AI\_system\}$ intrinsically classifies $AI\_system$ as a member of the class $Keyword$, while there is no rule to establish the type of the individual $fairness$.  
 
\item \textit{Semantic equivalence} is, in general terms, a key mechanism underlying the Semantic Interoperability model~\cite{heiler1995semantic}. It is critical also in this specific context to effectively foster a semantic convergence among the different frameworks. On a negative side, partial matching may introduce inaccuracies and, potentially, some uncertainty~\cite{pileggi2019web}. Semantic equivalences may be established at a different level in OWL. In the current version of the ontology, relatively simple equivalences are established among classes within the subset \textit{Materialisation \& Association} (Table~\ref{tab:classes}), where applications, scenarios and use cases are different concepts that may be considered semantically equivalent in context though. More complex relationships may be established between concepts of the same type or even of different types. Two examples are shown in Fig.~\ref{fig:equivalenceExamples}. In the example on the left side (Fig.~\ref{fig:homogeneousEquivalence}), the principle \textit{Fairness}, which is defined in both input frameworks, is consolidated by establishing a semantic equivalence between the two specifications. The other example (Fig.~\ref{fig:heterogeneousEquivalence}) provides a unified view of the concept \textit{Accountability}, which is a principle in one of the input frameworks and a requirement in the other one.

\begin{figure}[!h]
\centering
\begin{subfigure}{.48\textwidth}
\centering
    \includegraphics[width=\textwidth]{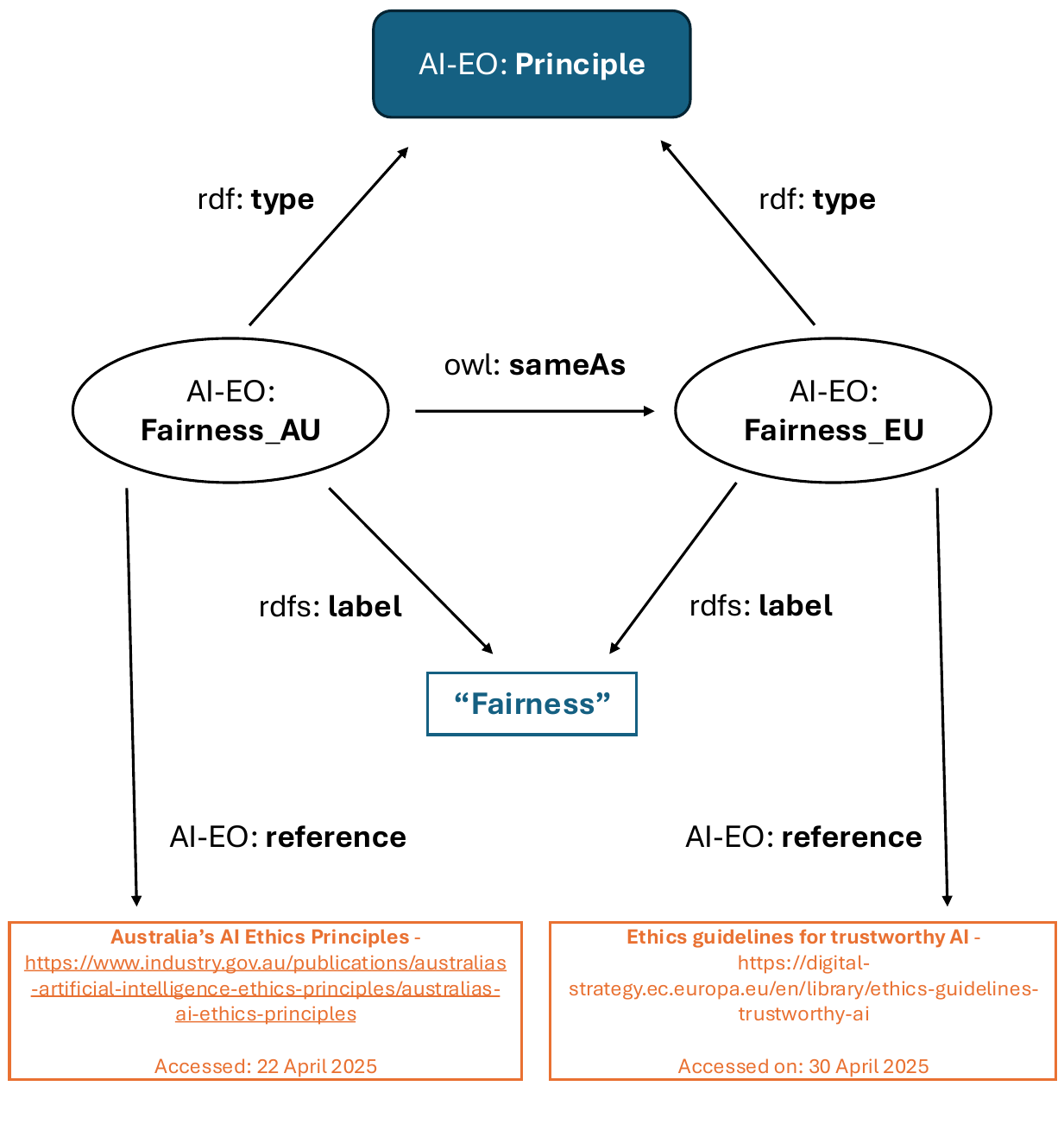}
    \caption{Equivalence of two principles.}\label{fig:homogeneousEquivalence}
\end{subfigure}
~
\begin{subfigure}{.48\textwidth}
\centering
    \includegraphics[width=\textwidth]{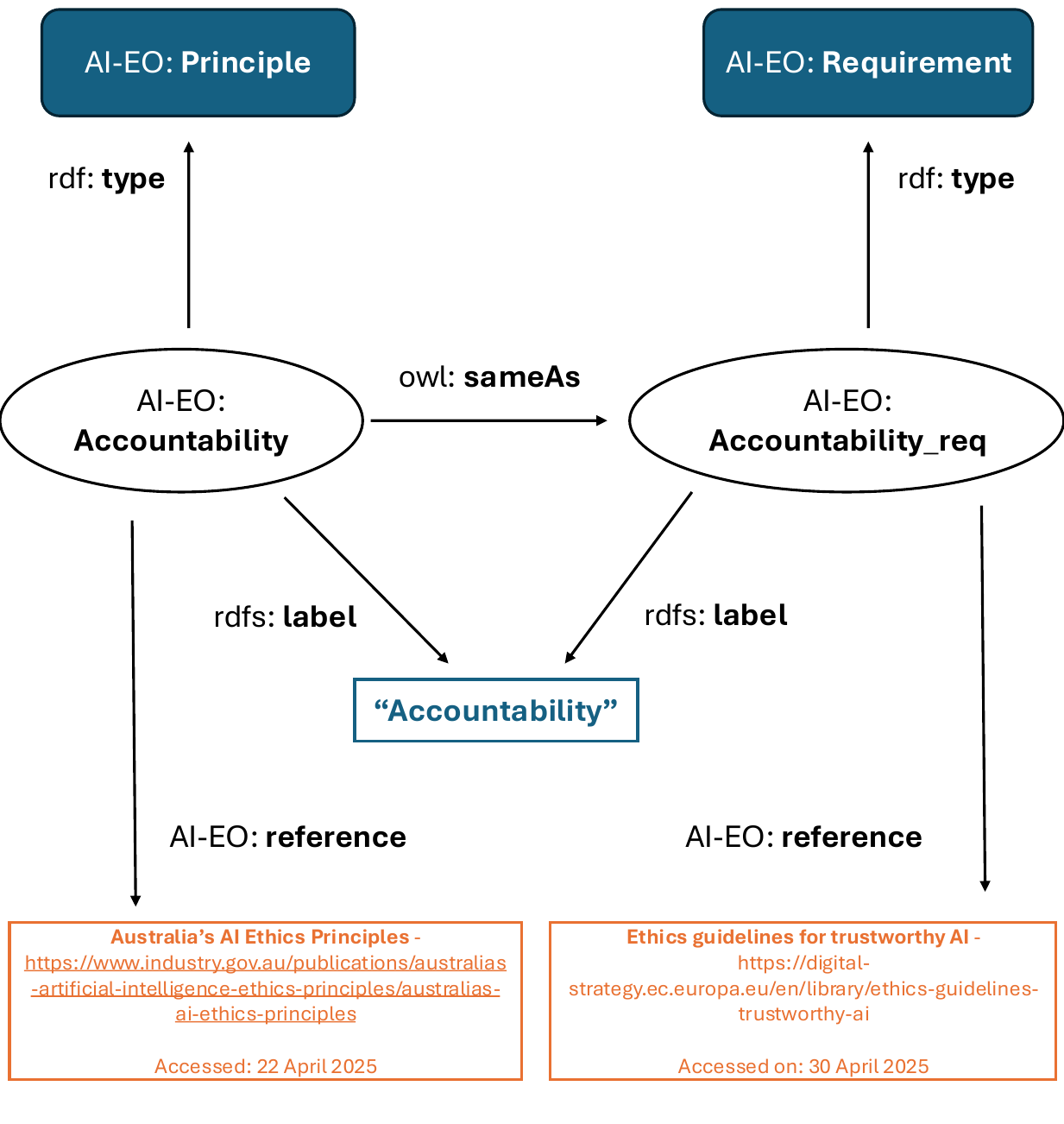}
    \caption{Equivalence of a principle and a requirement.}\label{fig:heterogeneousEquivalence}
\end{subfigure}
\caption{Examples of semantic equivalence involving concepts of the same type and of a different type.}\label{fig:equivalenceExamples}
\end{figure}

\end{itemize}

\subsection{Visualization \& Applications}

Ontologies are data infrastructure which aim at modelling knowledge in a given domain. This theoretical application-agnostic focus is often integrated with a certain degree of customisation or specialisation to better fit concrete applications.

The very first value provided by AI-EO is the capability to perform complex query on a federated view. An ontology-based system may easily and directly generate an answer to very specific questions, such as: \textit{what principles are available across the different frameworks? How is the same principle described in the different frameworks? What are example scenarios related to a given principle? What's the unique value of each framework?}

\begin{figure}[!h]
\centering
\fbox{\includegraphics[width=\textwidth]{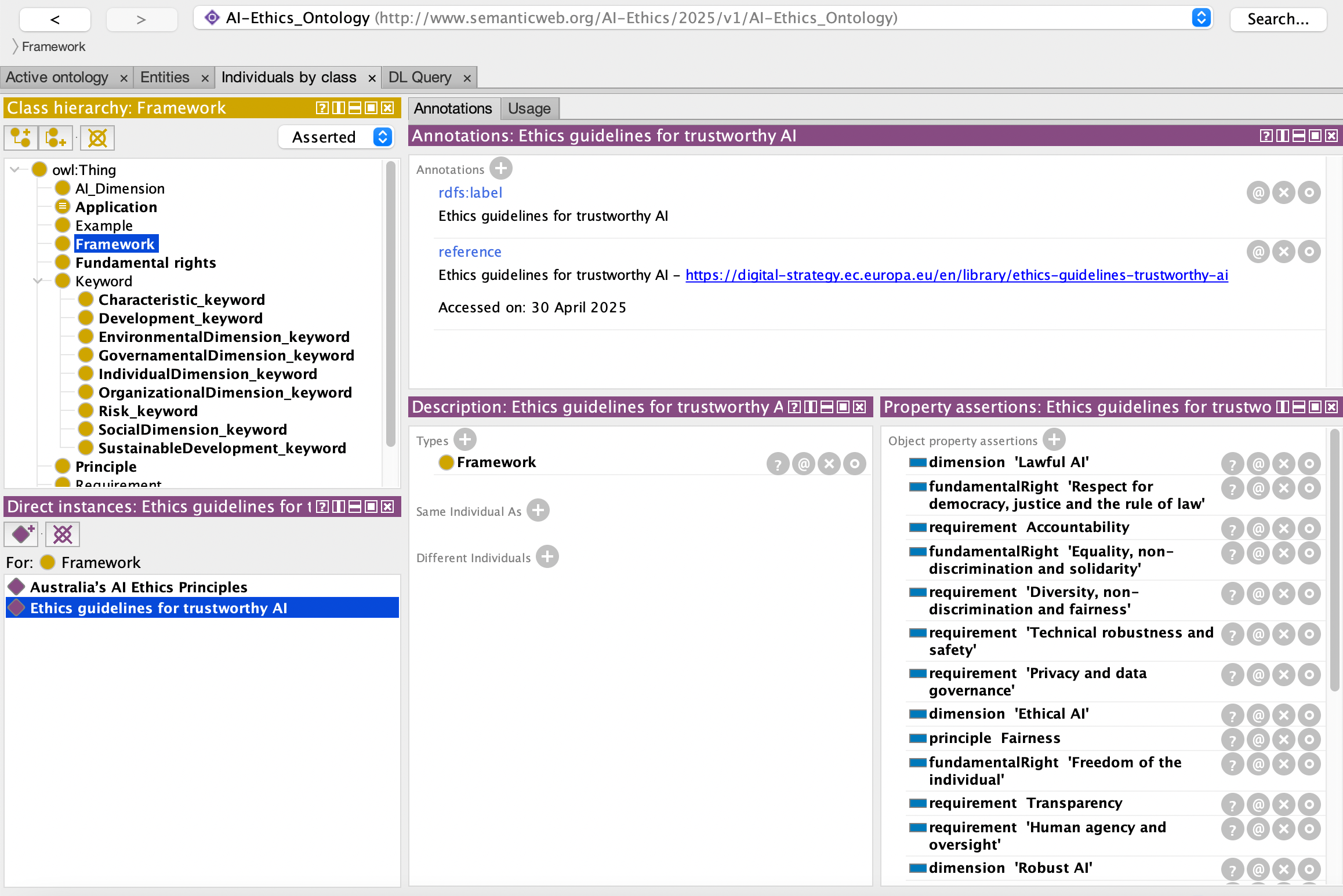}}
\caption{Visualization of the Ontology in Protege.}\label{fig:Protege}
\end{figure}

In more generic terms, the ontology intrinsically enables further knowledge building though dynamic incremental value, especially considering the potential support of developing tools such as Protege~\cite{gennari2003evolution}, which is explicitly designed to foster abstractions and usability looking at an extended range of potential users. A visualization of the current implementation in such tool is proposed in Fig.~\ref{fig:Protege}. 

\begin{figure}[!h]
\centering
\begin{subfigure}{\textwidth}
\centering
    \includegraphics[width=.8\textwidth]{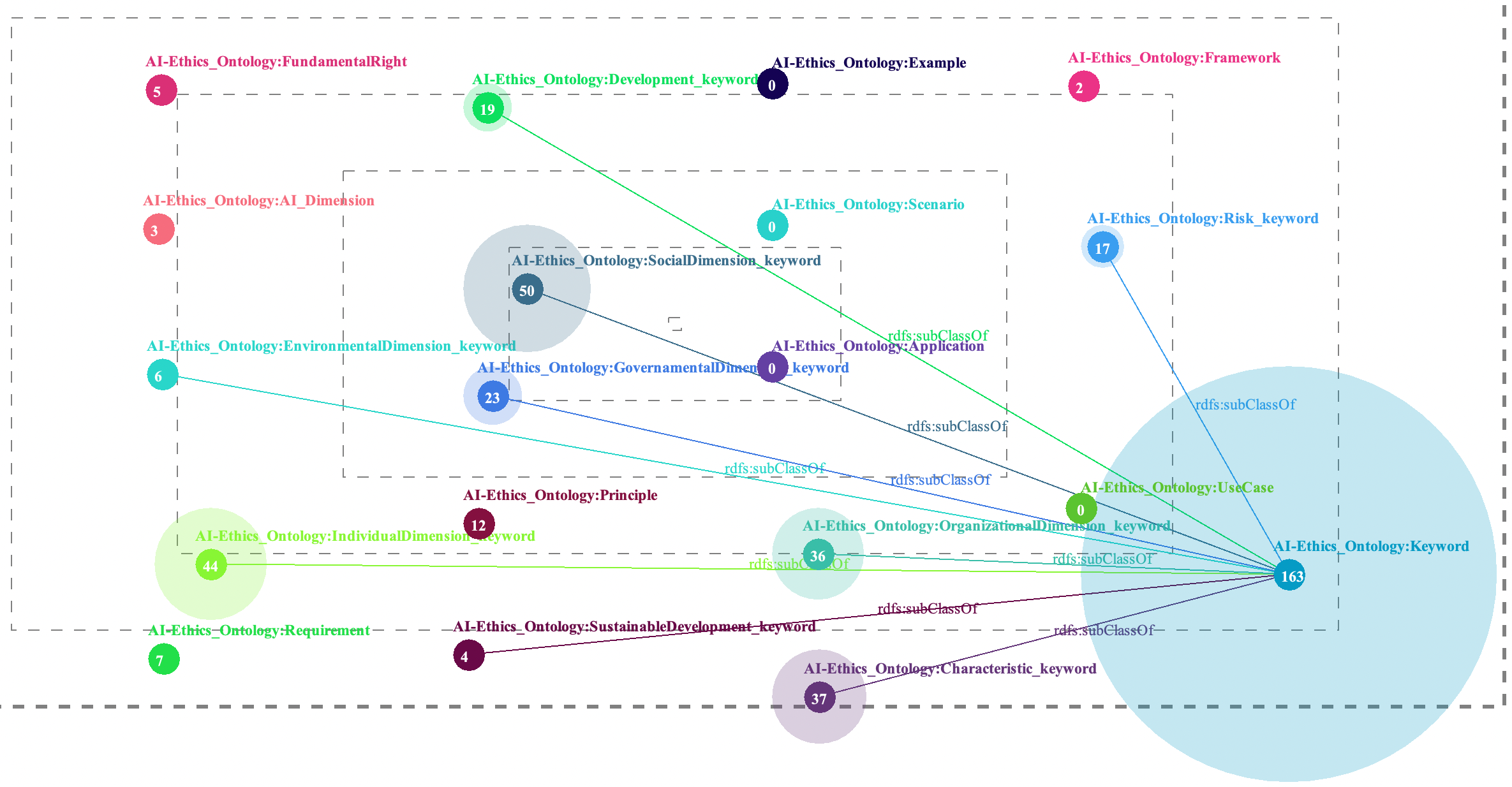}
    \caption{Classes, summary of associated individuals and class hierarchy.}
\end{subfigure}
\hfill
\begin{subfigure}{\textwidth}
\centering
    \includegraphics[width=.8\textwidth]{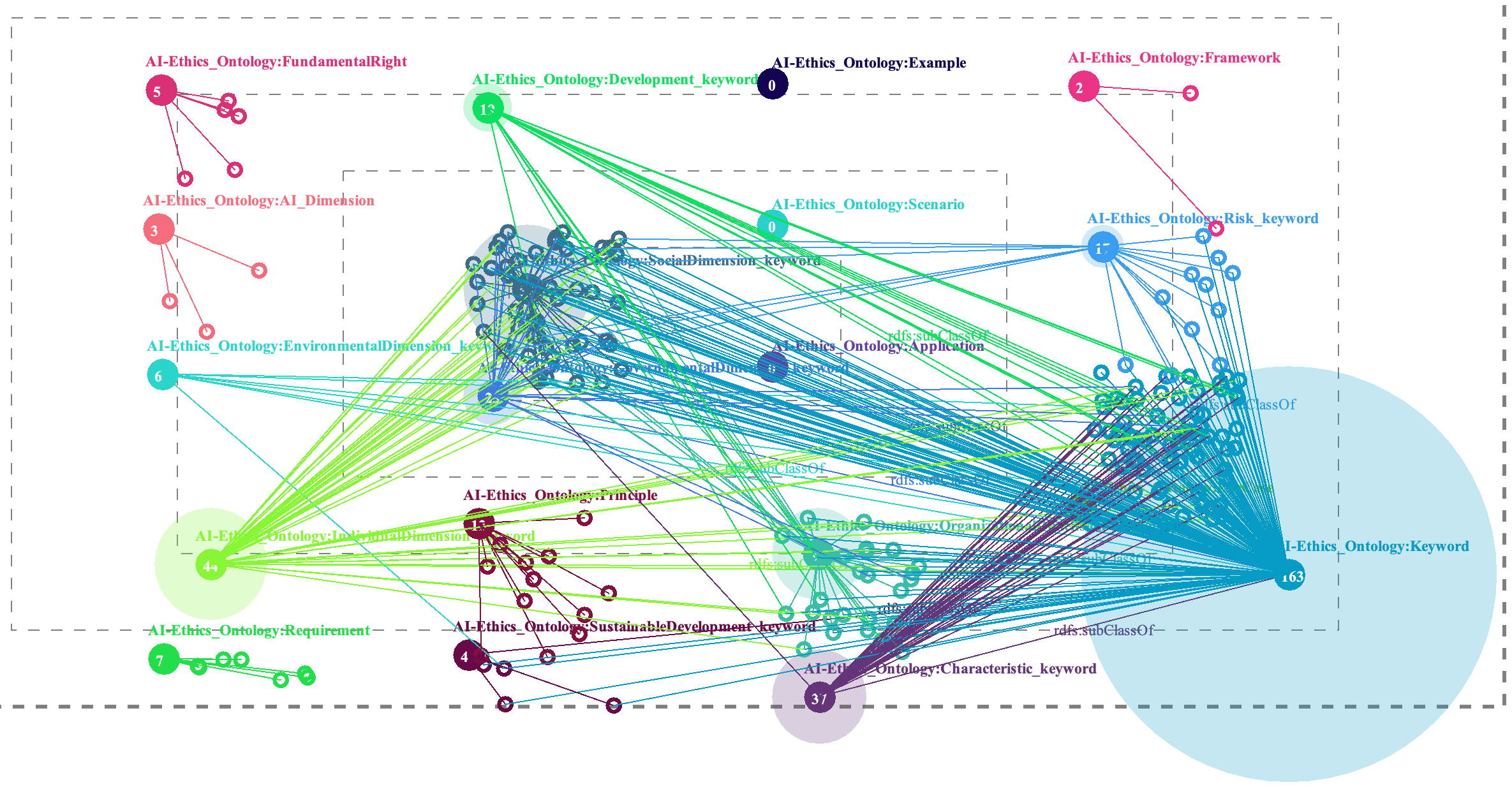}
    \caption{Addition of an explicit visualisation of individuals.}
\end{subfigure}
\hfill
\begin{subfigure}{\textwidth}
\centering
    \includegraphics[width=.8\textwidth]{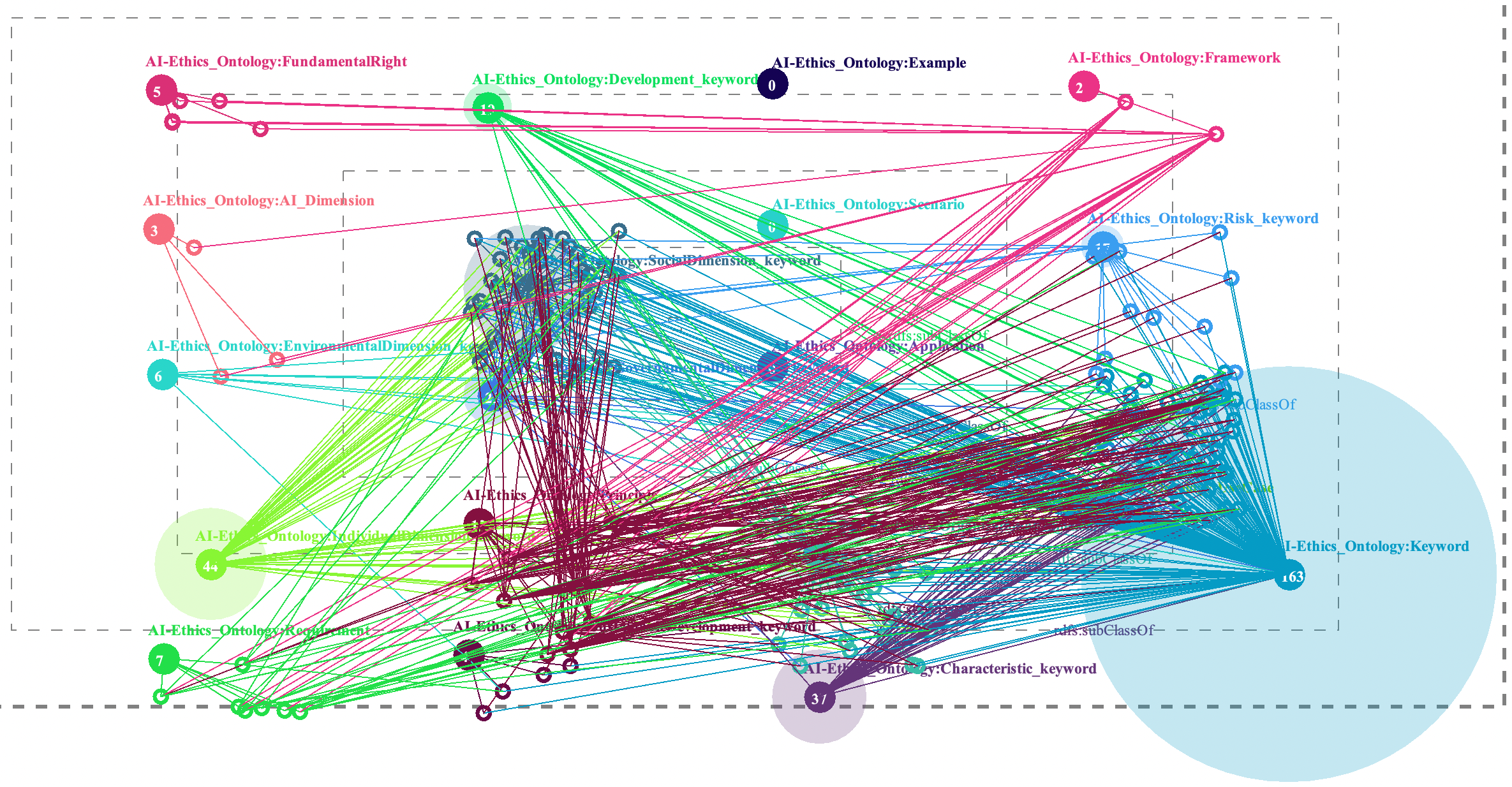}
    \caption{Extended view including also the relationships among instances.}
\end{subfigure}
        
\caption{Visualization of the Ontology as a Knowledge Graph, assuming a different level of detail~\cite{pileggi2022getting}.}
\label{fig:KG}
\end{figure}

Additionally, the ontological approach is functional to support intelligible elaborations of knowledge, for instance though Knowledge Graphs~\cite{hogan2021knowledge,ji2021survey}, which are indeed often underpinned by ontological data~\cite{pileggi2022getting}. A visualization of the ontology as a Knowledge Graph is proposed in Fig.~\ref{fig:KG}.

Last but not least, the underlying design philosophy is expected to be very close to final users as based on application-level abstractions. That should allow its smooth integration and exploitation  as a self-contained knowledge asset within the different tools and applications (e.g. compliance checking).

\section{Current limitations and future development}\label{sec:discussion}

As previously discussed, the current development may be considered a relatively mature research prototype which provides a consistent proof-of-concept by enabling a dynamic knowledge building process as a response to evolving needs.

Current limitations and related future work may be summarised as follows: 

\begin{itemize}

\item Link to external vocabularies, including other ontologies in the same domain and in contiguous domains, as well as to more generic vocabularies. This consolidation within the Semantic Web may further stimulate interoperability and data interchange, potentially at a global scale.

\item The current development limited to two iterations aimed at proof-of-concept, rather than at a full development. This approach is more functional to the actual needs, as well as to constant change and evolution. Additionally, it allows a more effective management of the inherent complexity through progressive consolidation.

\item In terms of Knowledge Engineering, the development of the ontology has been performed in a relatively traditional way, with key human input and support from AI technology. The extraction and consolidation of the knowledge (for instance of key terminology and related hierarchy) is expected to be further structured in order to maximise and exploit the value provided by AI-powered tools in a context of scalable hybrid intelligence.

\item While tested within popular environments and research tools from the research community, the current prototype has not yet been externally validated. Such a validation is expected to be progressively enabled by applications, which should provide contextual valuable feedback in addition to further development iterations.  

\end{itemize}

\section{Conclusions}

By leveraging Semantic Technologies on the Web infrastructure and ontology-based knowledge representations, AI-Ethics Ontology (AI-EO) provides an abstracted semantic infrastructure to foster the convergence, interoperability and operationalization of the different frameworks for Trustworthy AI. 

The current implementation (v1.0) is freely available and results from the analysis of two relevant case studies to establish a dynamic development process in fact, as well as to enable its iterative evolution according to a formally-defined methodology. This approach is more functional to the actual needs, as well as to constant change and evolution. Additionally, it allows a more effective management of the inherent complexity through progressive consolidation.

The ontology provides a federated view that has been designed to be conceptually close to target applications, in a context of interoperability and adaptability as a natural response to change and usability. Last but not least, the ontological approach naturally fosters a descriptive over a prescriptive approach.
The current development may be considered a relatively mature research prototype, which is is expected to grow in concept and scale as a response to external validation by application.


\begin{thebibliography}{10}

\bibitem{acharya2025agentic}
Deepak~Bhaskar Acharya, Karthigeyan Kuppan, and B~Divya.
\newblock Agentic ai: Autonomous intelligence for complex goals--a
  comprehensive survey.
\newblock {\em IEEE Access}, 2025.

\bibitem{angles2008expressive}
Renzo Angles and Claudio Gutierrez.
\newblock The expressive power of sparql.
\newblock In {\em International Semantic Web Conference}, pages 114--129.
  Springer, 2008.

\bibitem{AU_principels}
Science Australian Government Department~of Industry and Resources.
\newblock {Australia’s AI Ethics Principles}.
\newblock
  https://www.industry.gov.au/publications/australias-artificial-intelligence-ethics-principles/australias-ai-ethics-principles.
\newblock Accessed: 12 May 2025.

\bibitem{bengio2024managing}
Yoshua Bengio, Geoffrey Hinton, Andrew Yao, Dawn Song, Pieter Abbeel, Trevor
  Darrell, Yuval~Noah Harari, Ya-Qin Zhang, Lan Xue, Shai Shalev-Shwartz,
  et~al.
\newblock Managing extreme ai risks amid rapid progress.
\newblock {\em Science}, 384(6698):842--845, 2024.

\bibitem{chatila2017ieee}
Raja Chatila, Kay Firth-Butterflied, John~C Havens, and Konstantinos
  Karachalios.
\newblock The ieee global initiative for ethical considerations in artificial
  intelligence and autonomous systems.
\newblock {\em IEEE Robotics and Automation Magazine}, 24(1):110, 2017.

\bibitem{chowdhury2023unlocking}
Soumyadeb Chowdhury, Prasanta Dey, Sian Joel-Edgar, Sudeshna Bhattacharya,
  Oscar Rodriguez-Espindola, Amelie Abadie, and Linh Truong.
\newblock Unlocking the value of artificial intelligence in human resource
  management through ai capability framework.
\newblock {\em Human resource management review}, 33(1):100899, 2023.

\bibitem{EU_act}
{European Commission}.
\newblock {AI Act}.
\newblock
  https://digital-strategy.ec.europa.eu/en/policies/regulatory-framework-ai.
\newblock Accessed: 3 June 2025.

\bibitem{EU_trustwortyAI}
{European Commission}.
\newblock {Ethics guidelines for trustworthy AI}.
\newblock
  https://digital-strategy.ec.europa.eu/en/library/ethics-guidelines-trustworthy-ai.
\newblock Accessed: 3 June 2025.

\bibitem{fei2022towards}
Nanyi Fei, Zhiwu Lu, Yizhao Gao, Guoxing Yang, Yuqi Huo, Jingyuan Wen, Haoyu
  Lu, Ruihua Song, Xin Gao, Tao Xiang, et~al.
\newblock Towards artificial general intelligence via a multimodal foundation
  model.
\newblock {\em Nature Communications}, 13(1):3094, 2022.

\bibitem{fui2023generative}
Fiona Fui-Hoon~Nah, Ruilin Zheng, Jingyuan Cai, Keng Siau, and Langtao Chen.
\newblock Generative ai and chatgpt: Applications, challenges, and ai-human
  collaboration, 2023.

\bibitem{gennari2003evolution}
John~H Gennari, Mark~A Musen, Ray~W Fergerson, William~E Grosso, Monica
  Crub{\'e}zy, Henrik Eriksson, Natalya~F Noy, and Samson~W Tu.
\newblock The evolution of prot{\'e}g{\'e}: an environment for knowledge-based
  systems development.
\newblock {\em International Journal of Human-computer studies}, 58(1):89--123,
  2003.

\bibitem{glimm2014hermit}
Birte Glimm, Ian Horrocks, Boris Motik, Giorgos Stoilos, and Zhe Wang.
\newblock Hermit: an owl 2 reasoner.
\newblock {\em Journal of automated reasoning}, 53:245--269, 2014.

\bibitem{golpayegani2022airo}
Delaram Golpayegani, Harshvardhan~J Pandit, and Dave Lewis.
\newblock Airo: an ontology for representing ai risks based on the proposed eu
  ai act and iso risk management standards.
\newblock In {\em Towards a Knowledge-Aware AI}, pages 51--65. IOS Press, 2022.

\bibitem{guarino2009ontology}
Nicola Guarino, Daniel Oberle, and Steffen Staab.
\newblock What is an ontology?
\newblock {\em Handbook on ontologies}, pages 1--17, 2009.

\bibitem{guizzardi2023ontology}
Renata Guizzardi, Glenda Amaral, Giancarlo Guizzardi, and John Mylopoulos.
\newblock An ontology-based approach to engineering ethicality requirements.
\newblock {\em Software and Systems Modeling}, 22(6):1897--1923, 2023.

\bibitem{guizzardi2024using}
Renata Guizzardi, Glenda Amaral, Giancarlo Guizzardi, and John Mylopoulos.
\newblock Using i* to analyze ethicality requirements following ontology-based
  requirements engineering.
\newblock In {\em Social Modeling Using the i* Framework: Essays in Honour of
  Eric Yu}, pages 183--204. Springer, 2024.

\bibitem{harrison2021ontology}
Andrew Harrison, Dayana Spagnuelo, and Ilaria Tiddi.
\newblock An ontology for ethical ai principles.
\newblock {\em Semantic Web Journal}, 2021.

\bibitem{heiler1995semantic}
Sandra Heiler.
\newblock Semantic interoperability.
\newblock {\em ACM Computing Surveys (CSUR)}, 27(2):271--273, 1995.

\bibitem{hernandez2024ontology}
Julio Hernandez, Delaram Golpayegani, and David Lewis.
\newblock Ontology-based approach for mapping concepts and requirements from
  regulations and standards: The case of the eu ai act and international
  standards.
\newblock In {\em Legal Knowledge and Information Systems}, pages 295--300. IOS
  Press, 2024.

\bibitem{hitzler2009owl}
Pascal Hitzler, Markus Kr{\"o}tzsch, Bijan Parsia, Peter~F Patel-Schneider,
  Sebastian Rudolph, et~al.
\newblock Owl 2 web ontology language primer.
\newblock {\em W3C recommendation}, 27(1):123, 2009.

\bibitem{hogan2021knowledge}
Aidan Hogan, Eva Blomqvist, Michael Cochez, Claudia d’Amato, Gerard~De Melo,
  Claudio Gutierrez, Sabrina Kirrane, Jos{\'e} Emilio~Labra Gayo, Roberto
  Navigli, Sebastian Neumaier, et~al.
\newblock Knowledge graphs.
\newblock {\em ACM Computing Surveys (Csur)}, 54(4):1--37, 2021.

\bibitem{houghtaling2023standardizing}
Michael~A Houghtaling, Sandro~Rama Fiorini, Nicola Fabiano, Paulo~JS
  Gon{\c{c}}alves, Ozlem Ulgen, Tam{\'a}s Haidegger, Joel~Lu{\'\i}s Carbonera,
  Joanna~Isabelle Olszewska, Brian Page, Zvikomborero Murahwi, et~al.
\newblock Standardizing an ontology for ethically aligned robotic and
  autonomous systems.
\newblock {\em IEEE Transactions on Systems, Man, and Cybernetics: Systems},
  54(3):1791--1804, 2023.

\bibitem{ji2021survey}
Shaoxiong Ji, Shirui Pan, Erik Cambria, Pekka Marttinen, and Philip~S Yu.
\newblock A survey on knowledge graphs: Representation, acquisition, and
  applications.
\newblock {\em IEEE transactions on neural networks and learning systems},
  33(2):494--514, 2021.

\bibitem{jones1998methodologies}
Dean Jones, Trevor Bench-Capon, and Pepijn Visser.
\newblock Methodologies for ontology development.
\newblock 1998.

\bibitem{kotis2006human}
Konstantinos Kotis and George~A Vouros.
\newblock Human-centered ontology engineering: The hcome methodology.
\newblock {\em Knowledge and Information Systems}, 10:109--131, 2006.

\bibitem{lassila2001semantic}
Ora Lassila, J~Hendler, and T~Berners-Lee.
\newblock The semantic web.
\newblock {\em Scientific american}, 284(5):34--43, 2001.

\bibitem{lewis2021ontology}
Dave Lewis, David Filip, and Harshvardhan~J Pandit.
\newblock An ontology for standardising trustworthy ai.
\newblock {\em Factoring Ethics in Technology, Policy Making, Regulation and
  AI}, 65, 2021.

\bibitem{li2023trustworthy}
Bo~Li, Peng Qi, Bo~Liu, Shuai Di, Jingen Liu, Jiquan Pei, Jinfeng Yi, and Bowen
  Zhou.
\newblock Trustworthy ai: From principles to practices.
\newblock {\em ACM Computing Surveys}, 55(9):1--46, 2023.

\bibitem{liang2022advances}
Weixin Liang, Girmaw~Abebe Tadesse, Daniel Ho, Li~Fei-Fei, Matei Zaharia,
  Ce~Zhang, and James Zou.
\newblock Advances, challenges and opportunities in creating data for
  trustworthy ai.
\newblock {\em Nature Machine Intelligence}, 4(8):669--677, 2022.

\bibitem{mikalef2021artificial}
Patrick Mikalef and Manjul Gupta.
\newblock Artificial intelligence capability: Conceptualization, measurement
  calibration, and empirical study on its impact on organizational creativity
  and firm performance.
\newblock {\em Information \& management}, 58(3):103434, 2021.

\bibitem{OECD}
OECD.AI.
\newblock {OECD AI Principles overview}.
\newblock https://oecd.ai/en/ai-principles.
\newblock Accessed: 19 May 2025.

\bibitem{patel2021present}
Archana Patel and Sarika Jain.
\newblock Present and future of semantic web technologies: a research
  statement.
\newblock {\em International Journal of Computers and Applications},
  43(5):413--422, 2021.

\bibitem{peroni2016simplified}
Silvio Peroni.
\newblock A simplified agile methodology for ontology development.
\newblock In {\em International Experiences and Directions Workshop on OWL},
  pages 55--69. Springer, 2016.

\bibitem{pileggi2019web}
Salvatore~F Pileggi.
\newblock Web of similarity.
\newblock {\em Journal of Computational Science}, 36:100578, 2019.

\bibitem{pileggi2022getting}
Salvatore~Flavio Pileggi.
\newblock Getting formal ontologies closer to final users through knowledge
  graph visualization: interpretation and misinterpretation.
\newblock In {\em International Conference on Computational Science}, pages
  611--622. Springer, 2022.

\bibitem{prestes2021first}
Edson Prestes, Michael~A Houghtaling, Paulo~JS Gon{\c{c}}alves, Nicola Fabiano,
  Ozlem Ulgen, Sandro~Rama Fiorini, Zvikomborero Murahwi, Joanna~Isabelle
  Olszewska, and Tam{\'a}s Haidegger.
\newblock The first global ontological standard for ethically driven robotics
  and automation systems [standards].
\newblock {\em IEEE Robotics \& Automation Magazine}, 28(4):120--124, 2021.

\bibitem{rismani2023does}
Shalaleh Rismani and AJung Moon.
\newblock What does it mean to be a responsible ai practitioner: An ontology of
  roles and skills.
\newblock In {\em Proceedings of the 2023 AAAI/ACM Conference on AI, Ethics,
  and Society}, pages 584--595, 2023.

\bibitem{sadovski2024towards}
Eran Sadovski, Itzhak Aviv, and Irit Hadar.
\newblock Towards a comprehensive ontology for requirements engineering for
  ai-powered systems.
\newblock In {\em International Working Conference on Requirements Engineering:
  Foundation for Software Quality}, pages 219--230. Springer, 2024.

\bibitem{shao2022tracing}
Zhou Shao, Ruoyan Zhao, Sha Yuan, Ming Ding, and Yongli Wang.
\newblock Tracing the evolution of ai in the past decade and forecasting the
  emerging trends.
\newblock {\em Expert Systems with Applications}, 209:118221, 2022.

\bibitem{simperl2014collaborative}
Elena Simperl and Markus Luczak-R{\"o}sch.
\newblock Collaborative ontology engineering: a survey.
\newblock {\em The Knowledge Engineering Review}, 29(1):101--131, 2014.

\bibitem{sirin2007pellet}
Evren Sirin, Bijan Parsia, Bernardo~Cuenca Grau, Aditya Kalyanpur, and Yarden
  Katz.
\newblock Pellet: A practical owl-dl reasoner.
\newblock {\em Journal of Web Semantics}, 5(2):51--53, 2007.

\bibitem{spyns2002data}
Peter Spyns, Robert Meersman, and Mustafa Jarrar.
\newblock Data modelling versus ontology engineering.
\newblock {\em ACM SIGMod Record}, 31(4):12--17, 2002.

\bibitem{suarez2011introduction}
Mari~Carmen Su{\'a}rez-Figueroa, Asunci{\'o}n G{\'o}mez-P{\'e}rez, Enrico
  Motta, and Aldo Gangemi.
\newblock Introduction: Ontology engineering in a networked world.
\newblock In {\em Ontology engineering in a networked world}, pages 1--6.
  Springer, 2011.

\bibitem{tudorache2020ontology}
Tania Tudorache.
\newblock Ontology engineering: Current state, challenges, and future
  directions.
\newblock {\em Semantic Web}, 11(1):125--138, 2020.

\bibitem{van2023dynamics}
Colin Van~Noordt and Luca Tangi.
\newblock The dynamics of ai capability and its influence on public value
  creation of ai within public administration.
\newblock {\em Government Information Quarterly}, 40(4):101860, 2023.

\bibitem{W3CStandards}
W3C.
\newblock {Semantic Web Standards}.
\newblock https://www.w3.org/2001/sw/wiki/Main\_Page.
\newblock Accessed: 19 May 2025.

\bibitem{RDFS}
W3C.
\newblock {RDF Schema 1.1}.
\newblock https://www.w3.org/TR/rdf-schema/, 2014.
\newblock Accessed: 12 May 2025.

\bibitem{yeung2020recommendation}
Karen Yeung.
\newblock Recommendation of the council on artificial intelligence (oecd).
\newblock {\em International legal materials}, 59(1):27--34, 2020.

\end{thebibliography}

\section*{Declarations}
\begin{itemize}
\item \textit{This research did not receive any specific grant from funding agencies in the public, commercial, or not-for-profit sectors.}
\item \textit{The authors declare that they have no known competing financial interests or personal relationships that could have appeared to influence the work reported in this paper.}
\item \textit{Data availability: the ontology implementation described in the paper (v1.0) is available at: \url{https://github.com/sfpileggi/AI-EO}.}
\end{itemize}

\end{document}